\documentclass[11pt]{article}
\usepackage{moriond,epsfig}

\def\be{\begin{equation}}
\def\ee{\end{equation}}
\def\bea{\begin{eqnarray}}
\def\eea{\end{eqnarray}}

\def\cO#1{{\cal{O}}\left(#1\right)}
\def\cF{{\cal{F}}}    %   the multiple emission correction factor

\begin{document}

\begin{flushright}
IPPP-03-25\\
PCPT-03-50\\
hep-ph/0305163\\
\end{flushright}

\vspace*{3cm}
\title{AUTOMATED RESUMMATION OF FINAL STATE OBSERVABLES IN QCD}

\author{G.\ ZANDERIGHI}

\address{IPPP and University of Durham, Department of Physics, Durham
  DH1 3LE, England}

\maketitle\abstracts{We present an innovative method to resum infrared
  and collinear logarithms appearing in distributions of jet
  observables in QCD. The method, based on a general master formula
  with applicability conditions, allows resummations at
  next-to-leading logarithmic accuracy in an automated way. As a
  sample application we present resummed results in hadronic dijet
  events.}
\section{Jet observables in QCD}
A special aspect of QCD in the Standard Model is that the regime where
the theory is strongly coupled is within reach of current experiments,
and indeed unavoidable.  This gives rise to a number of non trivial
phenomena, for instance one encounters non-convergent perturbative
(PT) expansions and large non-perturbative (NP) corrections even at
scales which are, at least formally, within the domain of perturbation
theory.  One of the main aims of today's QCD studies is the
comprehension of all these phenomena.  Apart from its fundamental
interest, a detailed QCD description is vital for any precision/search
physics in modern colliders.

The study of jet observables, event-shapes and jet-rates, is
particularly informative in that these quantities are calculable in
perturbation theory with a remarkably high accuracy, but they are
still sensitive to NP, low energy physics.  For brevity, in the
following I will refer to event-shapes only; these are infrared and
collinear safe observables describing the topology of the final state
in high energy collisions.

In recent years their mean values and distributions have been used to
measure the coupling constant $\alpha_s$,\cite{Bethke} the colour
factors \cite{SU3} and to study their hadronisation
corrections.\cite{NPexp}
The most discriminatory studies make use of distributions.  Integrated
distributions $\Sigma(v)$ are defined by requiring that the value of
the observable $V(k_1,\dots,k_n)$ -- a function of all $n$ momenta --
be less than a fixed value $v$
\begin{equation}
\label{eq:sigma}
  \Sigma(v) = \int dV \frac{1}{\sigma}\frac{d\sigma}{dV} \, 
\theta(v-V(k_1,\dots,k_n))\>. 
\end{equation}
These distributions are characterised by two kinematical scales, the
hard scale $Q$ of the process (for instance the c.o.m. energy in
$e^+e^-$ collisions, the virtuality of the gauge boson in DIS
processes, etc.)  and an observable scale $v\,Q$ whose magnitude is
determined by the typical momenta of secondary emissions.
In the more exclusive kinematical region $v\ll 1$ there is a strong
ordering between these two scales, so that the distribution
Eq.~\ref{eq:sigma} contains (large) logarithms of the ratio of the two
scales $L\equiv \ln(1/v)$.  They compensate the smallness of
$\alpha_s$, and to make predictions one needs to resum logarithmically
enhanced (next-to-leading NLL) terms to all orders, while in the
region where $v=\cO{1}$ fixed order (next-to-leading NLO) predictions
are reliable.

It is interesting to compare today's effort required to obtain NLO and
NLL predictions.
To compute distributions at a given PT order one needs to know the
emission probabilities (the squared matrix elements) at that
perturbative order. Given the QCD Lagrangian, these quantities are
calculable in principle at any order, though technically it becomes
extremely difficult to go beyond NLO.
Since PT matrix elements are independent of the observable, they can
be implemented in a general way in Fixed Order Monte
Carlos.\cite{FOMC} These are then widely exploited for experimental
studies, the only additional input needed being the definition of the
specific observable in the form of a computer routine.
Obtaining NLL descriptions also requires the knowledge of the emission
probabilities. These quantities mix different logarithmic orders, so
that one needs to reorganise the PT series so as to account for ({\em
  resum}) all leading (LL) $\exp\{\alpha_s^n L^{n+1}\}$ and
next-to-leading logarithmic (NLL) $\exp\{\alpha_s^n L^{n}\}$ terms.
Also, since one is resumming terms to all PT orders, one needs to know
the value of the observable given an arbitrary number of secondary
emissions.
These are the reasons why, so far, the description of each observable
has required a separate analytical calculation,\cite{NG} this limits
the experimental study of resummed predictions, as compared to fixed
order ones.
\section{Automated resummation}
The aim of this project is to automate the NLL resummation of
jet observables, so as to ease their experimental analysis.
To achieve this goal one has to understand the origin of all NLL
enhanced terms in observable distributions.  This reduces to two
steps: one needs to know the behaviour of the observable when only one
soft-collinear (SC) gluon is present in the final state (single
emission properties); furthermore one needs to understand the way in
which all emissions coherently determine the value of the observable
(multiple emission properties).
\subsection{Single emission properties}
To describe the behaviour of the observable due to a single emission
we fix an (arbitrary) Born event and simulate the emission of one soft
gluon collinear to each hard parton (leg) $\ell$.  We assume that the
observable can be parameterised as
\begin{equation}
\label{eq:vsimple}
 V(k)\simeq d_{\ell}\left(\frac{k_t}{Q}\right)^{a_\ell}
e^{-b_{\ell}\eta} g_\ell(\phi)\>, 
\end{equation}
where $k_t$, $\eta$ denote respectively the transverse momentum and
the rapidity with respect to the emitting hard leg and $\phi$ is the
azimuthal angle with respect to a plane.  Our numerical code verifies
automatically the assumed parameterisation and determines the
coefficients $a_\ell, b_\ell, d_\ell$ and the functions
$g_\ell(\phi)$.  This knowledge allows us to compute all the NLL terms
which account for hard collinear emissions, soft large angle radiation
and inclusive gluon splittings (running of the coupling). This gives
rise to a single (or simple) distribution $\Sigma_s(v) =e^{-R_s(v)}$,
where $R_s(v)$ is a LL Sudakov exponent (complete formulae can be
found elsewhere).\cite{autolet+autopaper} It also happens, by
construction, that $\Sigma_s$ is the full NLL resummed distribution
for a simpler observable $V_s$ whose value is determined only by the
largest emission
\begin{equation}
  \label{eq:v-vs}
V_s(k_1,\dots,k_n) = \max[V(k_1),\dots, V(k_n)]\>.
\end{equation}
\subsection{Multiple emission properties}
To fully resum the observable at NLL one also needs to relate the value of the
observable to the momenta of all secondary emissions,
i.e. to understand the observable specific mismatch between the full
observable $V$ and its simplified variant $V_s$.
An analytical treatment of this requires a detailed insight in the
kinematics of the recoiling hard partons and, for jet-rates, of the
recombination procedure, but this effect can be computed numerically
in a general way~\cite{BSZ} and is given as a pure NLL function
$\cF(R')$, with $R'\equiv d R_s(v_s)/d\ln(1/v_s)$.
\subsection{The master formula and its applicability conditions}
The knowledge of the single and multiple emission properties is
combined in a master formula~\cite{autolet+autopaper}
\begin{equation}
  \label{eq:master}
  \Sigma(v) = \Sigma_s(v_s) \cF(R')\>.   
\end{equation}
The formula Eq.~\ref{eq:master} does of course {\em not} apply to all
final state observables in QCD and is accompanied by a list of
conditions the observable has to fulfil:\cite{autolet+autopaper} it
should be recursively infrared and collinear
safe;\cite{autolet+autopaper} should vanish in the Born limit; should
be continuously global;\cite{NG} should behave as assumed in
Eq.~\ref{eq:vsimple} when only one SC emission is present;
additionally it should, at the present stage, satisfy some minor
(technical) requirements.  While this might seem a long list
practically the limiting condition is the requirement of globalness,
while {\em all} other conditions are satisfied by {\em all}
observables resummed so far.  Most importantly an essential feature of
our code is the ability to verify all properties automatically and to
resum the observable only when the correctness of the result is
guaranteed at NLL accuracy.
\subsection{Sample output: first ever resummations in hadronic dijet
  production} 
\begin{figure}[t]
\begin{minipage}{0.5\textwidth}
    \epsfig{file=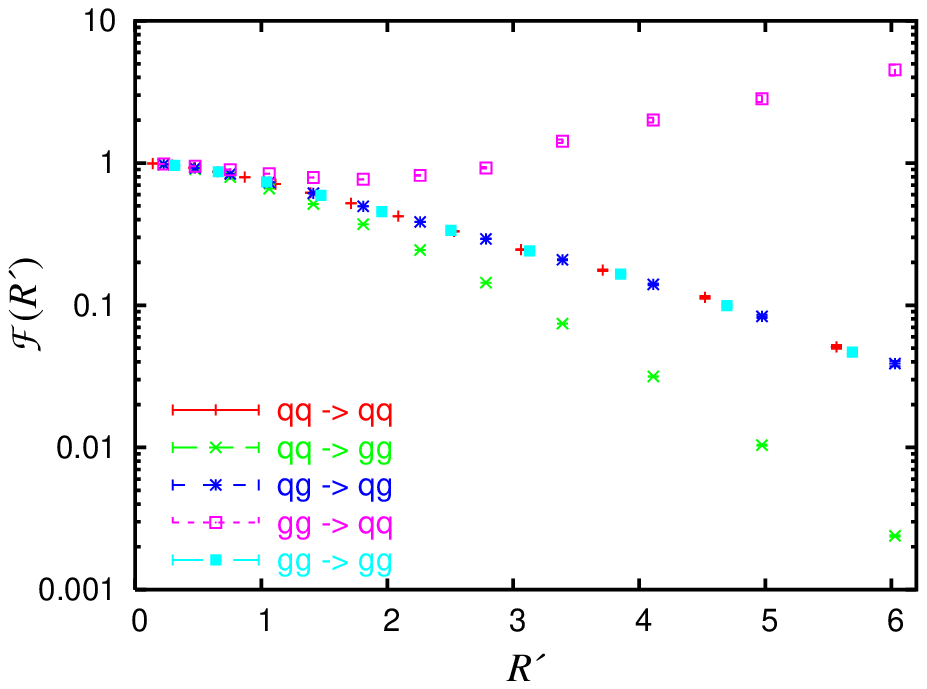, width=0.9\textwidth, angle=0}
\end{minipage}
\begin{minipage}{0.5\textwidth}
    \epsfig{file=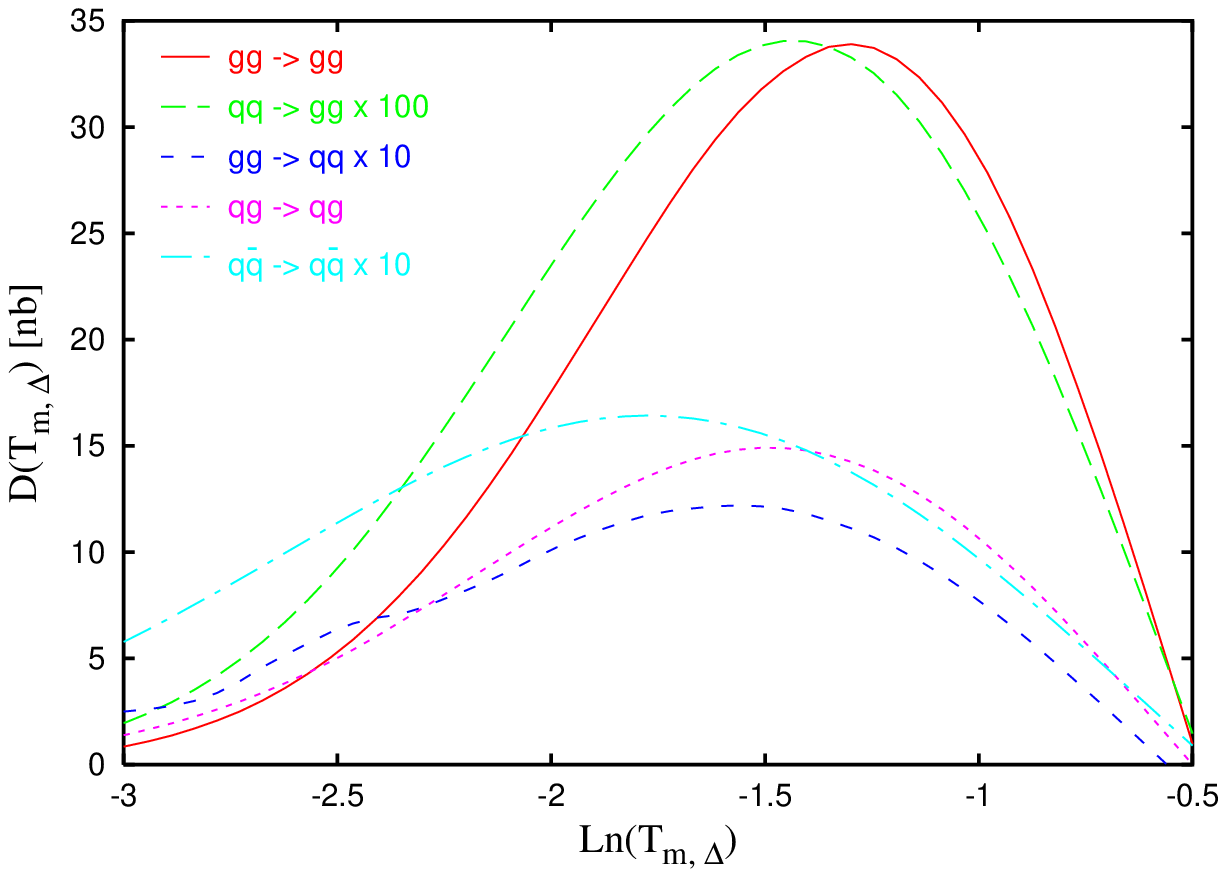, width=0.9\textwidth, angle=0}
\end{minipage}
  \caption{
    The function $\cF$ (left) and the NLL resummed distributions
    (right) for the indirectly global thrust
    minor.\label{fig:tmincut}}
\end{figure}

\begin{table}[t]
\begin{center}
\caption{Single emission properties for $T_{m, \Delta}$, 
%the indirectly global thrust minor, 
 $1$ and $2$ label the incoming legs, $3$ and $4$ the outgoing ones.}
\label{tab:tmincutprop}
\vspace{0.2cm}
 \begin{tabular}{| c | c | c | c | c | c |}
 \hline
 leg $\ell$ & $a_{\ell}$ & $b_{\ell}$ & $g_{\ell}(\phi)$ & $d_{\ell}$ & $ \langle \ln g_{\ell}(\phi) \rangle$  \\
 \hline
 \hline
1 & 1 & 0 & tabulated & 2 & -0.2201  \\ 
 \hline
2 & 1 & 0 & tabulated & 2 & -0.2201  \\ 
 \hline
3 & 1 & 0  & $\sin(\phi)$ & 2 & -$\ln(2)$\\ 
 \hline
4 & 1 & 0  & $\sin(\phi)$ & 2 & -$\ln(2)$\\ 
 \hline
 \end{tabular}
\end{center}
\end{table}
The most interesting application of our method is the resummation of
observables in hadronic dijet events.  As an example we present here
results for the indirectly global thrust minor~\footnote{Note that the
  second term guarantees the continuous globalness of the observable.}
\begin{equation}
T_{m, \Delta} \equiv \frac{\sum_{|\eta_i| < \Delta} |p_{xi}|}
{\sum_{|\eta_i| < \Delta} |\vec{p}_{ti}|}\, +\, 
  R_{t,\Delta}\>, \qquad   
R_{t,\Delta} \equiv \frac{\left|\sum_{|\eta_i| < \Delta} {\vec
      p}_{ti}\right|}{\sum_{|\eta_i| < \Delta} |\vec{p}_{ti}|}\,, 
\end{equation}
here $\vec{p}_{ti}$ denote the momenta transverse to the beam,
$p_{xi}$ the momenta out of the plane defined by the beam-axis and the
transverse thrust axis and $\Delta=\cO{1}$ denotes a rapidity cut.
The program verifies all applicability conditions; tabulates the
single emission properties, see Tab.~\ref{tab:tmincutprop}; computes
for all colour configurations the multiple emission function $\cF$,
see Fig.~\ref{fig:tmincut} (left); and exploits the master formula
Eq.~\ref{eq:master} to obtain the full NLL resummed prediction.
Fig.\ref{fig:tmincut} (right) shows differential distributions, for
several underlying hard subprocesses, for the Tevatron run II regime,
$\sqrt{s} = 1.96\,$GeV.  We select events having two outgoing jets
with $E_{t} > 50\,$GeV and $|\eta|<1$, fix $\Delta=1$, use the CTEQ6M
parton density set,\cite{CTEQ} corresponding to $\alpha_s(M_Z) =
0.118$ and set the factorisation and renormalisation scales $\mu_F$
and $\mu_R$ at the
partonic c.o.m. energy. 
\subsection{Final remarks, conclusions and outlook}
Results presented here have the quality of analytic NLL predictions,
though no human intervention is needed. This implies that any
hadronisation model can be applied, studies of renormalisation and
factorisation scale dependence can be carried out and, since the
answer is
free of spurious subleading  terms, a matching with
fixed order is feasible.

The first step in this project has been the numerical computation of
multiple-emission effects, which allowed us to resum three observables
in $e^+e^-$-collisions at a time.\cite{BSZ} We have now a code which
fully automates the resummation of a large class of jet observables;
as a result we obtained the first ever resummations in hadronic dijet
production.
As a next step we aim to automate the matching with fixed order
results. This will open up the possibility to carry out a vast amount
of phenomenological studies.
\section*{Acknowledgements}
I am enjoying working at this project with Andrea Banfi and Gavin
Salam.  A special thank to Yuri Dokshitzer and Pino Marchesini for
stimulating conversations and precious suggestions.

\end{document}